\documentstyle[12pt]{article}
\catcode`\@=11
  
\def\@normalsize{\@setsize\normalsize{15pt}\xiipt\@xiipt
\abovedisplayskip 14pt plus3pt minus3pt%
\belowdisplayskip \abovedisplayskip
\abovedisplayshortskip  \z@ plus3pt%
\belowdisplayshortskip  7pt plus3.5pt minus0pt}
\def\small{\@setsize\small{13.6pt}\xipt\@xipt
\abovedisplayskip 13pt plus3pt minus3pt%
\belowdisplayskip \abovedisplayskip
\abovedisplayshortskip  \z@ plus3pt%
\belowdisplayshortskip  7pt plus3.5pt minus0pt
\def\@listi{\parsep 4.5pt plus 2pt minus 1pt
            \itemsep \parsep
            \topsep 9pt plus 3pt minus 3pt}}
 
\def\underline#1{\relax\ifmmode\@@underline#1\else
        $\@@underline{\hbox{#1}}$\relax\fi}
\@twosidetrue
\relax

\catcode`@=12

\catcode`\@=11

\def\section{\@startsection{section}{1}{\z@}{3.5ex plus 1ex minus
   .2ex}{2.3ex plus .2ex}{\large\bf}}

\def\ps@headings{\def\@oddfoot{}\def\@evenfoot{}
\def\@oddhead{\hbox{}\hfill
        \makebox[.5\textwidth]{\raggedright\ignorespaces --\thepage{}--
        \hfill }}
\def\@evenhead{\@oddhead}
\def\subsectionmark##1{\markboth{##1}{}}
}
 
\ps@headings
 
\catcode`\@=12
 
\relax

\def\figcap{\section*{Figure Captions\markboth
        {FIGURECAPTIONS}{FIGURECAPTIONS}}\list
        {Fig. \arabic{enumi}:\hfill}{\settowidth\labelwidth{Fig. 999:}
        \leftmargin\labelwidth
        \advance\leftmargin\labelsep\usecounter{enumi}}}
 \relax
\def\tablecap{\section*{Table Captions\markboth
        {TABLECAPTIONS}{TABLECAPTIONS}}\list
        {Table \arabic{enumi}:\hfill}{\settowidth\labelwidth{Table 999:}
        \leftmargin\labelwidth
        \advance\leftmargin\labelsep\usecounter{enumi}}}
 \relax
\def\reflist{\section*{References\markboth
        {REFLIST}{REFLIST}}\list
        {[\arabic{enumi}]\hfill}{\settowidth\labelwidth{[999]}
        \leftmargin\labelwidth
        \advance\leftmargin\labelsep\usecounter{enumi}}}
 \relax
 
\catcode`\@=11
 
\def\marginnote#1{}
\newcount\hour
\newcount\minute
\newtoks\amorpm
\hour=\time\divide\hour by60
\minute=\time{\multiply\hour by60 \global\advance\minute by-
\hour}
\edef\standardtime{{\ifnum\hour<12 \global\amorpm={am}%
    \else\global\amorpm={pm}\advance\hour by-12 \fi
    \ifnum\hour=0 \hour=12 \fi
    \number\hour:\ifnum\minute<100\fi\number\minute\the\amorpm}}
\edef\militarytime{\number\hour:\ifnum\minute<100\fi\number\minute}

\def\draftlabel#1{{\@bsphack\if@filesw {\let\thepage\relax
  \xdef\@gtempa{\write\@auxout{\string
    \newlabel{#1}{{\@currentlabel}{\thepage}}}}}\@gtempa
    \if@nobreak \ifvmode\nobreak\fi\fi\fi\@esphack}
     \gdef\@eqnlabel{#1}}
\def\@eqnlabel{}
\def\@vacuum{}
\def\draftmarginnote#1{\marginpar{\raggedright\scriptsize\tt#1}}
\def\draft{\oddsidemargin -.5truein
        \def\@oddfoot{\sl preliminary draft \hfil
        \rm\thepage\hfil\sl\today\quad\militarytime}
        \let\@evenfoot\@oddfoot \overfullrule 3pt
        \let\label=\draftlabel
        \let\marginnote=\draftmarginnote
 
\def\@eqnnum{(\theequation)\rlap{\kern\marginparsep\tt\@eqnlabel}%
\global\let\@eqnlabel\@vacuum}  }
\def\preprint{\twocolumn\sloppy\flushbottom\parindent 1em
        \leftmargini 2em\leftmarginv .5em\leftmarginvi .5em
        \oddsidemargin -.5in    \evensidemargin -.5in
        \columnsep 15mm \footheight 0pt
        \textwidth 250mmin      \topmargin  -.4in
        \headheight 12pt \topskip .4in
        \textheight 175mm
        \footskip 0pt
 
\def\@oddhead{\thepage\hfil\addtocounter{page}{1}\thepage}
        \let\@evenhead\@oddhead \def\@oddfoot{} \def\@evenfoot{}
}
\def\titlepage{\@restonecolfalse\if@twocolumn\@restonecoltrue\onecolumn
     \else \newpage \fi \thispagestyle{empty}\c@page\z@
        \def\thefootnote{\fnsymbol{footnote}} }
\def\endtitlepage{\if@restonecol\twocolumn \else  \fi
        \def\thefootnote{\arabic{footnote}}
        \setcounter{footnote}{0}}  
\catcode`@=12
\relax
 
\def\ps@headings{\def\@oddfoot{}\def\@evenfoot{}
\def\@oddhead{\hbox{}\hfill
        \makebox[.5\textwidth]{\raggedright\ignorespaces --\thepage{}--
        \hfill }}
\def\@evenhead{\@oddhead}
\def\subsectionmark##1{\markboth{##1}{}}
}
 
\ps@headings
 
\relax
 
\def\firstpage#1#2#3#4#5#6{
\begin{document}
\newcommand{\newc}{\newcommand}
\newc{\ra}{\rightarrow}
\newc{\lra}{\leftrightarrow}
\newc{\beq}{\begin{equation}}
\newc{\eeq}{\end{equation}}
\newc{\bea}{\begin{eqnarray}}
\newc{\eea}{\end{eqnarray}}
\def\eps{\epsilon}
\def\la{\lambda}
\def\f{\frac}
\def\ne{\nu_{e}}
\def\nm{\nu_{\mu}}
\def\nt{\nu_{\tau}}
\def\sq{\sqrt{2}}
\def\ri{\rightarrow}
\newc{\sm}{Standard Model}
\newc{\smd}{Standard Model}
\newc{\barr}{\begin{eqnarray}}
 \newc{\earr}{\end{eqnarray}}
\begin{titlepage}
\nopagebreak
\title{\begin{flushright}
        \vspace*{-0.8in}
{ \normalsize  hep-ph/9711476 \\
ACT-17/97 \\
CERN-TH/97-336 \\
CTP-TAMU-47/97 \\
  IOA-16/1997 \\
November 1997 \\
}
\end{flushright}
\vfill
{#3}}
\author{\large #4 \\[1.0cm] #5}
\maketitle
\vskip -7mm
\nopagebreak
\begin{abstract}
{\noindent #6}
\end{abstract}
\vfill
\begin{flushleft}
\rule{16.1cm}{0.2mm}\\[-3mm]

\end{flushleft}
\thispagestyle{empty}
\end{titlepage}}
 
\def\simlt{\stackrel{<}{{}_\sim}}
\def\simgt{\stackrel{>}{{}_\sim}}
\date{}
\firstpage{3118}{IC/95/34}
{\large\bf Fermion Mass Textures in an $M$-Inspired
Flipped $SU(5)$ Model Derived from String}
{J. Ellis$^{\,a}$,  G.K. Leontaris$^{\,a,b}$,
S. Lola$^{\,a}$ and D.V. Nanopoulos$^{\,c,d,e}$}
{\normalsize\sl
$^a$CERN Theory Division, 1211 Geneva 23, Switzerland\\[2.5mm]
\normalsize\sl
$^b$Theoretical Physics Division, Ioannina University,
GR-45110 Ioannina, Greece\\[2.5mm]
\normalsize\sl
$^c$Center for Theoretical Physics, Department of Physics,\\[-1.0mm]
\normalsize\sl
  Texas A\&M
 University, College Station, TX 77843 4242,  USA\\[2.5mm]
\normalsize\sl
$^d$Astroparticle Physics Group, Houston Advanced Research Center (HARC),
\\[-1.0mm]
\normalsize\sl
The Mitchell Campus, Woodlands, TX 77381, USA\\[2.5mm]
\normalsize\sl
$^e$ Academy of Athens, Chair of Theoretical Physics, Division of Natural
Sciences,\\[-1.0mm]
\normalsize\sl
 28 Panepistimiou Ave., Athens GR-10679,  Greece.}
{We are inspired by the facts that $M$ theory may reconcile the
supersymmetric GUT scale with that of quantum gravity, and that 
it provides new avenues for low-energy supersymmetry breaking, to
re-examine a flipped $SU(5)$ model that has been derived from string and
may possess an elevation to a fully-fledged $M$-phenomenological model.
Using a complete analysis of all superpotential terms through the sixth order,
we explore in this model a new flat potential direction that provides
a pair of light Higgs doublets, yields
realistic textures for the fermion mass matrices, and
is free of $R$-violating interactions and dimension-five
proton decay operators.}
 
\newpage
\section{Introduction}

String model building is currently in a state of flux. Although
quasi-realistic string models have been
constructed~\cite{strimod}, particularly in the free-fermion formulation
of the
weakly-coupled heterotic string~\cite{sm}, further progress
was held back by poor understanding 
of the non-perturbative string effects that should determine the correct
string vacuum. Moreover, a major problem 
for most weakly-coupled string models was the apparent discrepancy between
the string unification scale calculated from first principles and the
supersymmetric grand 
unification scale inferred from measurements of the Standard Model
gauge couplings at LEP and elsewhere~\cite{scale}. In fact,
this discrepancy and a better understanding of 
non-perturbative string effects contributed to the motivation for 
study of the strong-coupling limit of string theory, and 
it is now
known that the unification-scale discrepancy may be removed if the
Theory of
Everything turns out to be a strong-coupling limit of $M$
theory~\cite{xxx},
corresponding to an eleventh dimension which is much larger than
the Planck length \cite{dimitri}. This may enable
string models constructed directly
in four dimensions~\cite{sm} to reconcile the measured
gauge couplings and four-dimensional Planck mass. 
Another issue which is cast in new light by advances in our
understanding of non-perturbative effects in string theory is
that of supersymmetry breaking. Previous upper limits on the
rank of the effective four-dimensional gauge group have been
abolished~\cite{candelas}, which may open new horizons for gaugino
condensation, and the Scherk-Schwarz mechanism has been
revived~\cite{Mtheor}.
Since it seems that many of the previous weakly-coupled string models
may be elevated to consistent vacua of $M$ theory, it is a good moment to
re-investigate them, and explore whether their remaining
phenomenological deficiencies can be overcome.

Among the phenomenological issues to be addressed by effective
low-energy models derived from strings before they can be
considered serious candidates for describing the elementary particle world
are the flat directions of the effective
potential and the associated choice of vacuum expectation values, 
the absence of light Higgs triplets and the presence of light electroweak
Higgs doublets, the texture of
the fermion mass matrices, $R$ violation and proton decay.
In particular, the stability of the proton had become
particularly troublesome with the advent of supersymmetric GUTs,
when it was realised that dimension-five operators of the
form $QQQL$ and/or $u^cu^cd^c e^c$ could be generated 
by the exchange of coloured GUT states, 
inducing rapid
proton decay~\cite{dim5}. In the string context, such operators could also
be induced by exchanges of heavy string modes. However, it has
recently been shown~\cite{efn} that there is
a class of free-fermion string models in which
an enhanced custodial symmetry forbids
the appearance of such dangerous operators to all orders
in perturbation theory, and arguments have been given~\cite{efn}
that these models may be elevated to $M$ theory. 

Since the problems of the unification scale and proton stability
may be on the way to resolution with such elevations of
`traditional' string models,
we now explore some of their other phenomenological aspects
in more detail, focussing in particular on predictions
for fermion masses and mixing angles.
There has been considerable work
on the fermion mass problem during the last few years. In order to
understand the
observed hierarchies and mixings in the most predictive way, model builders
have proposed specific textures of mass matrices at the unification
scale with a minimal number of parameters. 
Inspired by string model building, several groups have attempted
to use simple family U(1) or discrete symmetries to obtain the required
textures. However, despite valiant efforts~\cite{KN,af,akll},
a fully realistic pattern of fermion masses
and mixing has never been derived from a string model,
which is the objective of this paper.

We work in the context of three-generation superstring models derived in
the free-fermionic formulation~\cite{sm}, which
have the advantages of working directly in four
dimensions and yielding easily unified models that reduce to
the minimal supersymmetric extension of the Standard Model. 
The class of
models that we favour is based on string-derived versions
of flipped SU(5)~\cite{aehn}~\footnote{This
class of models corresponds geometrically to
$Z_2\times Z_2$ orbifold compactification at the maximally-symmetric point
in the Narain moduli space, and their three-generation nature
is directly related to the $Z_2\times Z_2$ orbifold structure.}. Using
a certain three-generation string basis in the free-fermion
formulation,
it is easy to produce variant models with similar group structures
but different phenomenological characteristics,
by making slight modifications
of the string basis. This gives us hope that minor `defects' of 
an otherwise successful model can be cured. 

Indeed, we exhibit in this paper an explicit string-derived
model with
a variant pattern of scalar vacuum expectation values that
ensures heavy Higgs triplets, provides two light Higgs doublets,
yields a qualitatively successful pattern of fermion masses and mixing,
and has neither $R$-violating interactions nor
proton decay operators to the order studied, which includes
all sixth-order terms in the effective superpotential.

\section{A Free-Fermion Model and its Spectrum}

{}From among the relatively rich variaty of
free-fermion models~\cite{aehn,alr,af2,LNY,lyk,oth}
we choose to work in the context of the $SU(5)\times U(1)$
model~\cite{aehn}.
This model is defined by a set of basis vectors defining
boundary conditions on the world-sheet fermions~\cite{sm}
that span a finite additive group, and the physical states in the Hilbert
space
of a given sector are obtained by 
acting on the vacuum with bosonic and fermionic 
operators and then applying generalized GSO projections that ensure
consistency with the string constraints. 
The construction of the flipped $SU(5)$ model can be seen in two
stages. First,
a set of five vectors $(1, S, b_{1,2,3})$ is introduced
which define an $SO(10)\times SO(6)\times E_8$ gauge group with N=1
supersymmetry. Next, adding the vectors
$b_{4,5},\alpha$\cite{aehn}~\footnote{These
correspond to Wilson lines in the orbifold formulation.},
the number of generations is reduced to three and the observable-sector 
gauge 
group obtained is $SU(5)\times U(1)$ accompanied by additional four $U(1)$ 
factors and a hidden-sector $SU(4)\times SO(10)$ gauge symmetry.

The massless spectrum generated by the above basis,
consists of the supergravity and gauge multiplets, the latter arising from
the Neveu-Schwarz sector,
and the seventy chiral superfields listed below with their
non-Abelian group representation contents and their
$U(1)$ charges. A generic feature shared with
all such $k=1$ constructions is that there are no adjoint or
higher-dimensional representations \cite{Drei,ECN}.
In this model, all states in
the observable sector belong to the  
1,5,10 of $SU(5)$ and their conjugates. This is why
higher-level constructions are needed to obtain traditional
GUT theories, such as
$SU(5)$ or $SO(10)$, which need adjoint Higgs
representations to break 
down to the Standard Model~\footnote{Important progress has recently
been made in higher-level string constructions of $SU(5)$ and $SO(10)$
models~\cite{AldIb}, which may eventually lead to more realistic versions.
Intrinsically $M$-theoretical compactifications might also be useful in
this respect.}.
On the other hand, flipped $SU(5)$ is one of the few GUTs
which only require Higgs representations smaller than the
adjoint, since
$10$ and $\overline{10}$ representations suffice to break the 
symmetry down to $SU(3)\times SU(2)\times U(1)$. We recall also that
the observable quarks and leptons are in the
$10,\bar{5},1$, but with assignments and electric charges
`flipped' relative to conventional $SU(5)$.

\begin{center}
{\bf Field Content of the Flipped $SU(5)$ String Model}
\end{center}

\begin{table}[h]
\centering
\begin{tabular}{|c||c||c|}
\hline
$F_1(10,\frac{1}{2},-\frac{1}{2},0,0,0)$ &
$\bar{f}_1(\bar{5},-\frac{3}{2},-\frac{1}{2},0,0,0)$ &
$\ell_1^c(1,\frac{5}{2},-\frac{1}{2},0,0,0)$ \\
 
$F_2(10,\frac{1}{2},0,-\frac{1}{2},0,0)$ &
$\bar{f}_2(\bar{5},-\frac{3}{2},0,-\frac{1}{2},0,0)$ &
$\ell_2^c(1,\frac{5}{2},0,-\frac{1}{2},0,0)$ \\
 
$F_3(10,\frac{1}{2},0,0,\frac{1}{2},-\frac{1}{2})$ &
$\bar{f}_3(\bar{5},-\frac{3}{2},0,0,\frac{1}{2},\frac{1}{2})$ &
$\ell_3^c(1,\frac{5}{2},0,0,\frac{1}{2},\frac{1}{2})$ \\
 
$F_4(10,\frac{1}{2},-\frac{1}{2},0,0,0)$ &
$f_4(5,\frac{3}{2},\frac{1}{2},0,0,0)$ &
$\bar\ell_4^c(1,-\frac{5}{2},\frac{1}{2},0,0,0)$ \\
 
$\bar{F}_5(\overline{10},-\frac{1}{2},0,\frac{1}{2},0,0)$ &
$\bar{f}_5(\bar{5},-\frac{3}{2},0,-\frac{1}{2},0,0)$ &
$\ell_5^c(1,\frac{5}{2},0,-\frac{1}{2},0,0)$ \\
\hline
\end{tabular}
\label{table:4}
 
\vspace*{0.5 cm}
 
\centering
\begin{tabular}{|c||c||c|}
\hline
$h_1(5,-1,1,0,0,0)$ & $h_2(5,-1,0,1,0,0)$ & $h_3(5,-1,0,0,1,0)$ \\
$h_{45}(5,-1,-\frac{1}{2},-\frac{1}{2},0,0)$ & & \\
\hline
\end{tabular}

\vspace*{0.5 cm}
 
\centering
\begin{tabular}{|c||c||c|}
\hline
$\phi_{45}(1,0,\frac{1}{2},\frac{1}{2},1,0) $ &
$\phi_{+}(1,0,\frac{1}{2},-\frac{1}{2},0,1) $ &
$\phi_{-}(1,0,\frac{1}{2},-\frac{1}{2},0,-1) $ \\
$\Phi_{23}(1,0,0,-1,1,0)$ &
$\Phi_{31}(1,0,1,0,-1,0)$  &
$\Phi_{12}(1,0,-1,1,0,0)$ \\
$\phi_i(1,0,\frac{1}{2},
-\frac{1}{2},0,0)$ &
$\Phi_i(1,0,0,0,0,0)$ & \\
\hline
\end{tabular}
 
\vspace*{0.5 cm}
 
\centering
\begin{tabular}{|c||c||c|}
\hline
$\Delta_1(0,1,6,0,-\frac{1}{2},\frac{1}{2},0)$ &
$\Delta_2(0,1,6,-\frac{1}{2},0,\frac{1}{2},0)$ &
$\Delta_3(0,1,6,-\frac{1}{2},-\frac{1}{2},0,
\frac{1}{2})$ \\
$\Delta_4(0,1,6,0,-\frac{1}{2},\frac{1}{2},0)$ &
$\Delta_5(0,1,6,\frac{1}{2},0,-\frac{1}{2},0)$ & \\
\hline
\end{tabular}

\vspace*{0.5 cm}
 
\centering
\begin{tabular}{|c||c||c|}
\hline
$T_1(0,10,1,0,-\frac{1}{2},\frac{1}{2},0)$ &
$T_2(0,10,1,-\frac{1}{2},0,\frac{1}{2},0)$ &
$T_3(0,10,1,-\frac{1}{2},-\frac{1}{2},0,\frac{1}{2})$ \\
$T_4(0,10,1,0,\frac{1}{2},-\frac{1}{2},0)$ &
$T_5(0,10,1,-\frac{1}{2},0,\frac{1}{2},0)$ & \\
\hline
\end{tabular}
\caption{
{\it The chiral superfields are listed with their
quantum numbers \cite{aehn}.
The $F_i$, $\bar{f}_i$, $\ell_i^c$,
as well as the
$h_i$, $h_{ij}$ fields and the singlets
are given in terms
of their
$ SU(5) \times U(1)' \times U(1)^4$ 
quantum numbers. 
Conjugate fields have opposite $U(1)' \times U(1)^4$
quantum numbers.
The fields $\Delta_i$ and $T_i$ are tabulated in terms
of their $U(1)' \times SO(10) \times SO(6) \times U(1)^4$
quantum numbers. }
}
\end{table}
 
As can be seen explicitly above, the matter and
Higgs fields in this string model carry additional charges under surplus
$U(1)$ symmetries~\cite{aehn}, there are a number
of neutral singlet fields, and a hidden-sector 
matter fields which transform
non-trivially under the $SU(4)\times SO(10)$ gauge symmetry,
as sextets under $SU(4)$, namely $\Delta_{1,2,3,4,5}$, and as
decuplets under $SO(10)$, namely $T_{1,2,3,4,5}$. There are also
fourplets of the $SU(4)$ hidden symmetry which possess fractional
charges, however, 
as we discuss later, these are confined and
will not concern us here.

We recall that the flavour assignments of the light
Standard Model particles in this model are as follows:
\bea
\bar{f}_1 : \bar{u}, \tau, \; \; \;
\bar{f}_2 : \bar{c}, e/ \mu, \; \; \;
\bar{f}_5 : \bar{t}, \mu / e \nonumber \\
F_2 : Q_2, \bar{s}, \; \; \;
F_3 : Q_1, \bar{d}, \; \; \;
F_4 : Q_3, \bar{b} \nonumber \\
\ell^c_1 : \bar{\tau}, \; \; \;
\ell^c_2 : \bar{e}, \; \; \;
\ell^c_5 : \bar{\mu}
\label{assignments}
\eea
up to mixing effects which we discuss later.

\section{An Interesting Flat Direction of the Effective Potential}

Any string model such as that reviewed above has a degenerate
potential with many flat directions along which combinations
of scalar fields may acquire large vacuum expectation values.
The detailed phenomenological properties of the model depend
on the choices of these moduli of the vacuum, which are
subject to flatness conditions associated with both the $D$
and the $F$ terms in the effective potential, but cannot be
fixed unambiguously with the string technology currently
available. A new feature of this paper is a different choice of
flat direction from those considered 
previously~\cite{KN,fldir}.
We choose non-zero vacuum expectation values for the 
following singlet and hidden-sector fields:
\beq
\Phi_{31}, \bar{\Phi}_{31}, \Phi_{23}, 
\bar\Phi_{23},\phi_2, \bar{\phi}_{3,4},  \phi^-, 
 \bar\phi^+ ,  \phi_{45},  \bar{\phi}_{45}, \Delta_{2,3,5}, T_{2,4,5}
\label{nzv}
\eeq
The vacuum expectation values of the hidden-sector fields
must satisfy additional constraints 
\beq
 T_{3,4,5}^2=T_i\cdot T_4=0,\,\, \Delta_{3,5}^2=0,\,\, T_2^2+\Delta_2^2=0
\label{hcon}
\eeq
which are imposed by the flatness conditions.
As we discuss below, an acceptable scenario for supersymmetry breaking
may still be possible, despite this breaking of the
hidden-sector gauge group, at least within the context of $M$-theory.

We now discuss in more detail the $F$-flatness conditions.
We have verified that
a pattern of vacuum expectation values of the form (\ref{nzv})
is compatible with flatness up to sixth order non-renormalizable terms 
in the superpotential. Here and later in the paper, we make
a computerized search of all possible superpotential terms up to this
order, discarding all terms that are disallowed by gauge
symmetries and string selection rules 
\cite{KLN,RT}. We do not calculate explicitly
the remaining terms, but assume that they appear with generic
coefficients of order unity. For the
vacuum expectation values of interest, the relevant $F$-flatness
conditions are:
 

\begin{eqnarray}
\frac{\partial W}{\partial \Phi_{12}} & : &
e^{i \gamma}
\Phi_{23}\Phi_{31} + \frac{1}{2} \phi_2^2 \approx  0
 \\
\frac{\partial W}{\partial T_4} & : &
\frac{1}{\sqrt{2}} \phi_2 T_5 +
T_4 (\Phi_{23} + \frac{1}{2} e^{i \delta_1}
\bar{\Phi}_{31} \phi_2^2) \approx   0
 \\
\frac{\partial W}{\partial T_5} & : &
\frac{1}{\sqrt{2}} \phi_2 T_4 +
T_5 (\Phi_{31} + \frac{1}{2} e^{i \delta_2}
\bar{\Phi}_{23} \phi_2^2) \approx  0
 \\
\frac{\partial W}{\partial \bar{\Phi}_{12}} & : &
e^{i w} \bar{\Phi}_{23} \bar{\Phi}_{31} +
\frac{1}{2} (\bar{\phi}_3^2 + \bar{\phi}_4^2)+(F_1\bar{F}_5)^2 \approx  0
 \\
\frac{\partial W}{\partial \phi_{2}} & : &
T_2\cdot  T_4 \Delta_2\cdot  \Delta_5 \approx  0
 \\
\frac{\partial W}{\partial \Delta_4} & : &
\frac{1}{\sqrt{2}} \Delta_5 \bar{\phi}_3 +
\Delta_4 {\bar{\Phi}_{23} + 2 e^{i p_1} \Phi_{31}
(\bar{\phi}_3^2 + \bar{\phi}_4^2 )}\approx  0
 \\
\frac{\partial W}{\partial \Delta_5} & : &
\frac{1}{\sqrt{2}} \Delta_4 \bar{\phi}_3 +
\Delta_5 {\bar{\phi}_{31} + 2 e^{i p_2} \Phi_{23}
(\bar{\phi}_3^2 + \bar{\phi}_4^2 )}\approx  0
 \\
\frac{\partial W}{\partial T_{2}} & : &
T_2 ( \Phi_{31} + 2 e^{i p_3} \phi_2^2 \bar{\Phi}_{23} )
+ \phi_2 \Delta_2 \Delta_5 T_4 \approx  0
 \\
\frac{\partial W}{\partial \Delta_{2}} & : &
\Delta_2 ( \Phi_{31} + 2 e^{i p4} \phi_2^2 \bar{\Phi}_{23} )
+ \phi_2 T_2 T_4 \Delta_5 \approx  0 \\
\frac{\partial W}{\partial \Phi_{3}} & : &
\phi_{45}\bar\phi_{45} + \phi_i\bar\phi_i\approx 0,\;\;\; i=4,\pm
\end{eqnarray}
where $\delta_i,w,p_i$ are phases that are in principle
calculable, that we leave indefinite. 
The approximate equality means that these conditions are
valid up to a certain order. One should bear in
mind that they will be further
modified when higher-order corrections are taken into account.

We see from the above that if we demand
$\Delta_2\cdot \Delta_5 \neq 0$, which is needed in order to obtain
non-zero (2,3) mixing in the $V_{CKM}$ matrix, as we discuss
later, we are forced to impose the condition $T_2\cdot T_4 = 0$.
Suppressing constants of order unity and various phases, the above
equations may be satisfied if the following relations hold
between the different non-zero vacuum expectation values:
 \begin{eqnarray}
\phi_2^2 & \sim & \Phi_{31}\bar{\Phi}_{23}+\cdots 
\label{s1}
\\
 1+\cdots & \sim & |(\Phi_{31}\bar\Phi_{31}-1)
                                      (\Phi_{23}\bar\Phi_{23}-1)| 
                                      \label{s2}
\\
\bar{\phi}_3^2 + \bar{\phi}_4^2 
& \sim & \bar\Phi_{23}\bar\Phi_{31}+\cdots
\label{s3}
\\
\bar{\phi}_{3}^2+\cdots &\sim  & 
          \bar\Phi_{23}\bar\Phi_{31}(1-\Phi_{31}\bar\Phi_{31})
(1-\Phi_{23}\bar\Phi_{23}) \\
\label{s4}
\phi_{45} {\bar \phi}_{45} &\sim & \phi_i {\bar \phi}_i + \cdots 
\label{s5}
\end{eqnarray}
where the dots stand for higher-order corrections to the $F$-flatness
conditions. Some such corrections may
be crucial in ensuring that all the singlet
fields may have non-zero vacuum expectation values in the perturbative
regime, i.e., $\langle \phi_i \rangle \le M_{String}/10$.
In particular, we assume that (\ref{s5}) is valid only when
such corrections are taken into account, which can be easily
satisfied when at least one of
the singlet fields $\phi_{45},\bar{\phi}_{45}$ develops a relatively
small vev. 
It can be easily checked that the above set (\ref{nzv})
of vacuum expectation values (vevs) satisfies
also the $D$-flatness conditions.

An additional constraint on the vacuum expectation values of the
hidden fields above is that they should be consistent with
the confinement of fractionally-charged states~\cite{ELNfrac}. We have
verified
that there is a pattern of values for the $\Delta_{i}$ vevs
which preserves unbroken an $SO(4)$ subgroup of the hidden-sector
$SU(4)$. Moreover, the residual unbroken gauge group is
asymptotically free, so that all fractionally-charged states
are confined, though at a lower energy scale than 
in previous versions of flipped $SU(5)$. 

We also note that
the hidden-sector $SO(10)$ breaks down to $SO(7)$,
which may still be sufficient to seed supersymmetry breaking
by gaugino condensation. We note, however, that this may not be
necessary when the model is elevated to $M$ theory. In this
case, the rank of the
gauge group may be enhanced, providing other gauge subgroups
that might lead to gaugino condensation. Moreover, it has been
suggested that the Scherk-Schwarz mechanism 
may operate in $M$ theory~\cite{Mtheor}, either supplementing or replacing
gaugino condensation in a realistic scenario for supersymmetry breaking.

\section{Light Higgs Doublets}

We now discuss the appearance of light Higgs doublets
and the large masses required for dangerous colored particles that
share common $SU(5)$ representations with the standard
model matter. Taking the latter issue first, and working to sixth order in
the superpotential, as
above, we  find the following mass terms
\bea
{\cal W}& \ra& \Phi_{31}h_3\bar{h}_1 +\bar{\Phi}_{31}h_1\bar{h}_3+
              \Phi_{23}h_3\bar{h}_2 +\bar{\Phi}_{23}h_2\bar{h}_3+
              \phi_{45}\bar{h}_3h_{45}+\bar\phi_{45}h_3\bar{h}_{45}
\nonumber\\
        &+& F_1F_1(h_1+h_2\phi_i^2+h_{45}\Phi_{31}\phi_{45}) \nonumber\\
        &+& \bar{F}_5\bar{F}_5(\bar{h}_2+
            \bar{h}_{45}\bar\phi_{45}\Phi_{23}+\bar{h}_1\phi_i^2
              +\Delta_{1,4}^2\bar{h}_3+T_1^2\bar{h}_1)
\label{tripl}
\eea
With our choice of vacuum expectation values, it can be easily checked
that
all  extra colour-triplet pairs become massive.

Turning now to the more delicate issue of the survival
of light Higgs doubles, we note that the
mass matrix for the Higgs fiveplets 
$h_{1,2,3}$, $h_{45}$ which may include the doublets 
required in the Standard Model takes the following form
to fifth order:
\begin{equation}
m_{h} =  \left(\begin{array}{cccc}
0  & \Phi_{12} & \bar\Phi_{31} & T_{5}^2\bar\phi_{45}  \\
\bar{\Phi}_{12} &   0   & \Phi_{23} &\Delta_{4}^2\bar{\phi}_{45} \\
\Phi_{31}  &  \bar\Phi_{23}& 0  &\bar\phi_{45} \\
\Delta_5^2 &T_4^2\phi_{45}&\phi_{45}&0
\end{array}\right), \label{higgs}
\end{equation}
With our choice (\ref{nzv}) of singlet vacuum expectation values:
$\langle \Phi_{12},\bar{\Phi}_{12}\rangle =0$
and with the supplementary conditions $\langle\Delta_i^2\rangle =0$
and $\langle T_i^2\rangle =0$, there are two
massless combinations
\cite{FLAT,KTJR} $h\sim \cos\theta h_1 +\cdots$, 
and $\bar{h}\sim \bar{h}_{45}+\cdots$.

Going on to sixth order, we find the following superpotential
terms that might {\it a priori} mix Higgs doublets and leptons:
\bea
{\cal W}&\ra& \bar F_5 F_2^2 \bar{f}_2\bar h_{45}+F_2\Delta_2\Delta_5\bar{f}_5\bar{h}_{45}
 + F_1T_1T_4\Phi_{2,4}\bar{f}_5\bar{h}_{45}\nonumber\\
&+&F_1\bar\phi_1\phi_{45}\Phi_{31}\bar{f}_1\bar{h}_2
+F_1\phi_{45}
(\bar\Phi_{12}\bar\phi_1+\phi_2\Phi_4)\bar f_1\bar h_3+
F_1\bar\phi_2\Phi_4\bar\Phi_{12}\bar f_1\bar h_{45}
\eea
However, these are not in fact dangerous.
The first two terms do not contribute to the Higgs mass matrix
because we choose $\langle F_2\rangle =0$. Moreover, the
choice (\ref{nzv}) also implies that
$\langle\Phi_i\rangle = \langle\bar\phi_{1,2}\rangle = 0$, so
the Higgs mass matrix (\ref{higgs}) remains intact through sixth order.
A complete dicussion of the light Higgs multiplets would need an
enumeration of many higher-order terms in the superpotential, which
would take us beyond the scope of this paper.

To the order studied, 
$\bar{h}$ (which contains 
a component of $\bar{h}_{45}$) remains light and hence is
available to give a mass
to the top quark through the coupling $F_4\bar{f}_5\bar{h}_{45}$.
Similarly, $h$ is available to give a mass to
the bottom quark, also via a trilinear coupling, whose
magnitude depends on a Higgs mixing angle $\theta$,
which is as yet undetermined. This model
therefore predicts a heavy top quark, but allows the bottom to be
significantly lighter, without requiring a large ratio of Higgs
vacuum expectation values.

\section{Fermion Mass Matrices and Mixing}

A deeper treatment of fermion mass matrices requires the
consideration of higher-order non-renormalizable terms which
fill in entries that vanish at lower order \cite{aehn}. Due to the
additional $U(1)$ symmetries in this type of model, 
and string selection rules \cite{KLN}, only a few Yukawa
couplings are available at any given order. A complete
discussion would require going to very high order, and would need,
for consistency, to discuss flat directions and the choice of
vacuum expectation values to comparable order. As before, we
restrict our attention to terms of at most sixth order, which are
sufficient to discuss interesting qualitative features of the fermion
mass matrices.

To this order, for $\langle h_2 \rangle \ll 
\langle h_1 \rangle $,
the {\bf down-quark} mass matrix takes the form
\begin{eqnarray}
M_D = \left(
\begin{array}{ccc}
0 & \Delta_2\Delta_3\bar\Phi_{23}& \Delta_5\Delta_3\bar\phi_{3} \\
\Delta_2\Delta_3\bar\Phi_{23}&
(\bar{\phi}_3^2 +
\bar{\phi}_4^2) 
&\Delta_2\Delta_5\bar{\phi}_4 \\
\Delta_5\Delta_3\bar\phi_{3} &\Delta_2\Delta_5\bar\phi_{4} & 1
\end{array}
\right)\lambda_b(M_{GUT}) \langle h_1\rangle
\label{down}
\end{eqnarray}
where at the unification scale
  $\lambda_b(M_{GUT})=\sqrt{2}g$. 
The zero entry should be understood as being of higher than
sixth order, and each of
the non-zero higher-order entries should be understood as
having numerical factor of order unity, with possible phases.
We notice that these entries have the right orders of magnitude
to reproduce the correct hierarchy of the down quark masses. 
In the approximation $\Delta_2\Delta_5\bar{\phi}_4 < 
\Delta_2\Delta_3\bar{\Phi}_{23}
\sim \Delta_2\Delta_3$, we find
a bottom-quark Yukawa coupling of magnitude $g \sqrt{2}$ to the
$h_1$ component of the $h$ field, so that
$m_b\sim g\sqrt{2} \hbox{sin} \theta \langle h\rangle$, which may be
much smaller than $m_t$, and the (1,2) mixing element
is given by
\beq
V_{12}^d = \sqrt{\frac{m_d}{m_s}} = 
\frac{  \Delta_2 \Delta_3 \bar{\Phi}_{23} }
{ \bar{\phi}_3^2 +  \bar{\phi}_4^2 }
\eeq

The {\bf up-quark} mass matrix in the particular flat
direction we study takes the form
\bea
M_{U} =
\left ( 
\begin{array}{ccc}
0 & 0 & \Delta_3 \Delta_5 \bar{\phi}_3 \\
0 & \bar{\phi}_4 & \Delta_2 \Delta_5 \bar{\phi}_4 \\
0 & \Delta_2 \Delta_5 & 1
\end{array}
\right ) \; \lambda_t(M_{GUT})\langle\bar{h}_{45}\rangle ,
\eea
with the same remarks as above concerning the meaning of the
zero entries and the presence of unspecified numerical
coefficients of order unity in the non-zero entries.
Since $m_u \neq 0$ only to higher order, we see that
$m_u<m_d$ is a natural possibility.
It is striking that the above string-derived Ansatz for the
quark mass matrices belongs to one~\cite{LN,LV} of the few cases~\cite{RRR}
based on the appearance of texture zeros which describe
correctly the low-energy quark masses and mixings using only a minimal set
of parameters at the GUT scale.
We should point out that the above matrices are defined at the unification
scale. To compare with the experimentally measured fermion masses, one
has to take into account the renormalisation group effects. 

We finally discuss the Cabibbo-Kobayashi-Maskawa (CKM) mixing
matrix. From the particular form of our  mass matrices, we deduce     
that the $(1-2)$ Cabibbo angle is essentially obtained from the down 
quark mass matrix. 
Ignoring for simplicity the $(1-3)$ mixing,
we find that $V^{CKM}_{12}\sim V_{12}^d$ approximately~\cite{KN}.
Specifically, after diagonalization of the
quark mass matrices, one finds the following form of the CKM
matrix to first order in perturbation theory 
and  up to order-unity coefficients
\begin{equation}
V^{CKM} \approx  \left(\begin{array}{ccc}
1 &  \Delta_2 \Delta_3 
\bar{\Phi}_{23} /
(\bar{\phi}_3^2 + \bar{\phi}_4^2 )
& \Delta_3 \Delta_5 \bar{\phi_3} 
\\ 
-\Delta_2 \Delta_3 
\bar{\Phi}_{23} /
(\bar{\phi}_3^2 + \bar{\phi}_4^2  )
& 1 & 
\Delta_2 \Delta_5 \bar{\phi_4}  \\
-\Delta_3   \Delta_5 \bar{\phi_3} 
 &  -\Delta_2 \Delta_5 \bar{\phi_4} 
 & 1
\end{array}\right), 
\label{CKM}
\end{equation}
where 
the vacuum expectation values of the fields are normalised with respect
to
the heavy mass scale of the theory, which should presumably be
identified with some $M$-theory scale $\sim 10^{16}$ GeV~\cite{xxx}.


{\bf Charged Leptons} in this model are accommodated in the $\bar{f_1},
\bar{f_2}$, $\bar{f}_5$ and
$\ell^c_1,\ell^c_2,\ell^c_5$ representations in the light spectrum shown
above. The remaining representations $f_{4},\bar{f}_3$,
$\bar{\ell}^c_4,\ell^c_3$ are expected to gain masses from terms
\begin{equation}
(f_{4}\bar{f}_3+ \bar{\ell}^c_4\ell^c_3) 
\left(T_3\cdot T_5(\phi_2\bar{\Phi}_{23}
  + \bar{\phi}_2\bar{\Phi}_{31})\right)
\end{equation}
Trilinear superpotential couplings yield
\begin{equation}
\bar{f}_1\ell^c_1 h_1 + (\bar{f}_2\ell^c_2  + \bar f_5\ell^c_5) h_2
\end{equation}
where one should expect as before higher-order corrections
involving products of the singlet and hidden-sector fields
$\phi_i,\phi^{\pm}$ and $\bar\phi_i,\bar\phi^{\pm}$, etc..
In order to obtain the known hierarchy of lepton masses, we see that
a sensible choice is to accommodate the $\tau$ as
$\ell^c_1,\bar{f_1}$,
while choosing $ \langle  h_1\rangle \gg  
\langle h_2\rangle $  for the
Higgs vacuum expectation values. Indeed, as can be seen directly 
from the superpotential, now bottom and $\tau$ couplings are exactly the
same at the unification scale \cite{CEG}, which agrees with the
measured values
after renormalisation effects are taken into account.

It is not clear yet what combinations
of $\bar{f}_{2,5}$ and $\ell^c_{2,5}$
 should be interpreted as first- and
second-generation leptons, because
there are higher-order terms mixing these latter fields
with some heavy states. We plan to return to these in a later
publication, together with the complicated issue of neutrino masses.


\section{$R$ Violation}

The types of terms discussed may not exhaust the pattern of
superpotential couplings in a string-derived model, since there
may, in particular, be terms that violate $R$ parity. These might
{\it a priori} cause serious phenomenological problems, because
there are stringent upper bounds on several individual couplings,
and especially on the simultaneous presence of some 
flavour-changing products of couplings~\cite{constraints}.
 
In our case, no renormalizable $R$-violating
contributions are allowed, because they are not invariant
under the GUT group, {\it in contrast} to conventional $SU(5)$. 
The flipped $SU(5)$ transformation properties of the $R$-violating
couplings are:
\bea
LL\bar{E} & \rightarrow & \bar{5} \times \bar{5} \times 1
\nonumber \\
LQ\bar{D} & \rightarrow & \bar{5} \times 10 \times 10
\nonumber \\
\bar{U}\bar{D}\bar{D} & \rightarrow &
\bar{5} \times 10 \times 10
\label{nonsinglet}
\eea
This could be regarded as another potential phenomenological
asset of flipped $SU(5)$, in the absence of any confirmed
$R$-violation in Nature. 

However, effective terms that
break $R$-parity may be generated by non-renormalizable
higher-order terms~\cite{KT,GR},
and in all the cases (\ref{nonsinglet}) we need 
simply to add a field
that transforms as a $10$ of $SU(5)$ in order to
obtain an $SU(5)$-invariant combination. If and when this field
gets a vacuum expectation value, effective operators 
may be generated. The only candidate for such a field is
$F_1$, but we must check in each case whether the
transformations of the fields match under
the $U(1)$ of the flipped $SU(5)$, as well as the other
$U(1)$ factors, and also check the string selection rules.

Working as before to sixth order in the superpotential, 
we find in our flat direction the following candidate
field combinations that are invariant under the gauge
symmetries of the theory:

\underline{ $\bar{U}\bar{D}\bar{D}$ operators :}
\bea
\bar{U}_1\bar{D}_2\bar{D}_3 F_1 \phi_{45} \Phi_{31} \nonumber
\eea

\underline{$ LQ\bar{D}$ operators :}
\bea
L_1Q_2\bar{D}_2 F_1 \phi_{45}\bar{\Phi}_{23}, \; \; \;
L_2Q_2\bar{D}_2 F_1 \phi_{45}\bar{\Phi}_{23} \nonumber \\ 
L_1Q_3\bar{D}_3 F_1 \phi_{45} \Phi_{31}, \; \; \;
L_2Q_3\bar{D}_3 F_1 \phi_{45} \Phi_{31} \nonumber \\
L_3Q_2\bar{D}_3 F_1 \phi_{45} \Phi_{31}, \; \; \;
L_3Q_3\bar{D}_2 F_1 \phi_{45} \Phi_{31} \nonumber
\eea

\underline{
$ LL\bar{E}$ operators :}
\bea
L_1L_2\bar{E}_1 F_1 \phi_{45}\bar{\Phi}_{23}, \; \; \;
L_1L_2\bar{E}_2 F_1 \phi_{45}\bar{\Phi}_{23} \nonumber \\
L_1L_3\bar{E}_3 F_1 \phi_{45} \Phi_{31}, \; \; \;
L_2L_3\bar{E}_3 F_1 \phi_{45} \Phi_{31} \nonumber
\eea
However, it turns out, by inspection, that none of these satisfies
all of the string selection rules!

At the present level of understanding, we cannot exclude the possibility
that $R$-violating couplings may appear at higher orders, and the
upper limits on some couplings and combinations are so severe that
it would be necessary for many higher-order terms to vanish before
any model could be considered safe. Nevertheless, we consider the
vanishing to this order a very positive point for this particular
string model.

\section{Dimensional-Five Proton-Decay Operators}

In this model, fourth-order superpotential terms provide
no dimension-five proton-decay operators.
The following are the potentially-dangerous dimension-five operators
generated by fifth-order non-renormalizable terms~\cite{elnpd,gl}:
\bea
F_4F_4F_3\bar{f}_3\Phi_{31},& F_2F_2F_3\bar{f}_3 \bar\Phi_{23},
 & F_1F_1F_3\bar{f}_3\Phi_{31}\\
F_3\bar{f}_3\bar{f}_1\ell^c_1\Phi_{31},&
 F_3\bar{f}_3\bar{f}_5\ell^c_5\bar{\Phi}_{23},&F_3\bar{f}_3\bar{f}_2\ell^c_2\bar\Phi_{23}\\
F_3\bar{f}_2\bar{f}_2\ell^c_3\Phi_{31},&
 F_3\bar{f}_1\bar{f}_1 \ell^c_3\Phi_{31},& F_3f_5\bar{f}_5\ell^c_3\bar\Phi_{23}
\eea
However, with our choice of vacuum expectation values (\ref{nzv}), none of
these terms are dangerous, because they do not involve particles
in the light Standard Model part of the spectrum.
At sixth order, the following potentially-dangerous operators appear:
\bea
F_4F_3F_3 \phi_+\bar f_5\bar\Phi_{23},&F_4F_3F_3\bar f_5 \bar\phi_-\Phi_{31}
\eea
However, the singlet fields $\phi_+$ and $\bar\phi_-$ do not acquire
vacuum expectation values (\ref{nzv}), so these also pose no problems.

In contrast to the string model described in~\cite{efn}, 
currently we do not see any symmetry reason why dimension-five
proton-decay operators should be absent in this model to all orders.
If they do appear at some higher order, this may signal that the
current model requires some adjustment, but this should be regarded
as an open question for the time being.
 
\section{Conclusions}

We have discussed in this paper a new variant of a flipped $SU(5)$
model derived from string~\cite{aehn} which has many positive features. It
exploits an appealing flat direction that accommodates a pair of light
Higgs doublets. These provide up- and down-quark mass matrices that
can reproduce the observed hierarchy of quark masses and $V_{CKM}$
matrix elements, since they have one of the patterns of flavour
texture zeros that had been proposed previously on purely
phenomenological grounds. The mass
matrices of charged leptons and neutrinos remain open issues, to
which we hope to return in a forthcoming paper. 
One very positive feature of this model is the absence of
$R$-violating interactions to the order studied. This is in
part due to the representation content of flipped $SU(5)$, but
is also due in part to string selection rules that forbid
certain couplings apparently permitted by the gauge symmetries.
Another positive feature of the model is the absence of
dimension-five proton-decay operators, which is essentially due to
details of the vacuum we have chosen and the corresponding
light-particle spectrum.

Outstanding problems that should be addressed in future work include
the discussion of lepton and neutrino masses mentioned above, and the
extension of all the analysis of superpotential terms to higher orders,
using systematically all the string selection rules as well as the gauge
symmetries of the model. In the longer run, we should also like to
explore the elevation of this model to an intrinsic $M$-theory
compactification. Here we have assumed that this is possible, and
only used the inspiration of $M$-theory to motivate the absence
of intermediate-scale degrees of freedom, which are no longer
needed to reconcile the bottom-up and top-down calculations of the
unification scale, and to motivate tolerance of a model with
symmetry breaking in the hidden sector, on the grounds that $M$ theory
may provide other mechanisms for supersymmetry breaking. Knowledge of
$M$-theory compactifications is accumulating, but none of those
exhibited explicitly so far has as many attractive features as the
model discussed here. Perhaps this or a related model may play a
useful r\^ole in focussing $M$-phenomenology?

{\bf Acknowledgements:  }
The work of S.L. is funded by a Marie Curie Fellowship
(TMR-ERBFMBICT-950565). The work of G.K.L. is  partially supported
by the TMR-contract ERBFMRX-CT96-0090. The work of D.V.N. is supported in part
by D.O.E. Grant DE-FG03-95-ER-40917.


\end{document}